\documentclass[aps,prl,twocolumn,groupedaddress,nofootinbib]{revtex4}
\usepackage[dvips]{graphicx}
\usepackage{amsfonts}
\usepackage{amssymb}
\usepackage{mathrsfs}
\usepackage{graphicx}
\usepackage{amsmath,amsthm,amssymb,amscd}
\usepackage{subfig}
\usepackage{hyperref}
\usepackage{multirow}
\usepackage{amsfonts}
\newif\ifdraft
\drafttrue
\newif\ifpreprint
\preprinttrue

\newcommand{\eq}{\begin{equation}}
\newcommand{\eqe}{\end{equation}}
\newcommand{\eqa}{\begin{eqnarray}}
\newcommand{\eqae}{\end{eqnarray}}

\newbox\charbox
\newbox\slabox
\def\s#1{{      
        \setbox\charbox=\hbox{$#1$}
        \setbox\slabox=\hbox{$/$}
        \dimen\charbox=\ht\slabox
        \advance\dimen\charbox by -\dp\slabox
        \advance\dimen\charbox by -\ht\charbox
        \advance\dimen\charbox by \dp\charbox
        \divide\dimen\charbox by 2
        \raise-\dimen\charbox\hbox to \wd\charbox{\hss/\hss}
        \llap{$#1$}
}}

\begin{document}

\title{
\ifpreprint
\hbox{\rm \small 
$\null$ \hskip 2 cm \hfill  \hskip 13.5 cm \hfill  PUPT-2462
} 
\fi
Gravitons, Bose-symmetry, and Bonus-scaling }
 
\author{David A. McGady and Laurentiu Rodina}
\email{\\dmcgady@princeton.edu \\ lrodina@princeton.edu}
\affiliation{Department of Physics, Princeton University, Princeton, NJ 08544}

\begin{abstract}

Modern on-shell S-matrix methods may dramatically improve our understanding of perturbative quantum gravity, but current foundations of on-shell techniques for General Relativity still rely on off-shell Feynman diagram analysis. Here, we complete the fully on-shell proof of Ref.~\cite{ST} that the recursion relations of Britto, Cachazo, Feng, and Witten (BCFW) apply to General Relativity tree amplitudes. We do so by showing that the surprising requirement of ``bonus" $z^{-2}$ scaling under a BCFW shift directly follows from Bose-symmetry. Moreover, we show that amplitudes in generic theories subjected to  BCFW deformations of identical particles necessarily scale as $z^{\rm even}$. When applied to the color ordered expansions of Yang-Mills, this directly implies the improved behavior under non-adjacent gluon shifts. Using the same analysis, three-dimensional gravity amplitudes scale as $z^{-4}$, compared to the $z^{-1}$ behavior for conformal Chern-Simons matter theory.

\end{abstract}

\maketitle


Mysteries abound at the interface between General Relativity and Quantum Field Theory. Particularly, graviton scattering amplitudes in maximally supersymmetric ${\cal N}=8$ Supergravity have surprisingly soft behavior in the deep ultraviolet (UV). To four loops, it has been shown that the critical dimension of supergravity is the same as ${\cal N}=4$ Super Yang-Mills, a conformally invariant theory free of UV divergences~\cite{Bern}. This result was obtained through the peculiar BCJ duality between color and kinematics, which relates graviton amplitudes to the squares of gluon amplitudes~\cite{BCJ}\cite{Bern:2010yg}. Other arguments, based the non-linearly realized $E_{7(7)}$ symmetry of ${\cal N}=8$ supergravity, predict UV finiteness to six-loops~\cite{HRM}. Yet others hint at a full finiteness (see e.g.~\cite{powercounting}).

Standard perturbative techniques, i.e. Feynman diagrams, lead to incredibly complicated expressions, and obfuscate general features of the theory. Reframing the discussion in terms of the modern analytic S-matrix has so far proven incredibly useful for discussing Yang-Mills theory (for example, in Ref. \cite{SYM}), and may provide crucial insights into quantum gravity as well. The on-shell program offers a different perspective on the principles of locality and unitarity, and their powerful consequences \cite{ST}\cite{beni}. It also provides a computational powerhouse, the BCFW on-shell recursion relation \cite{BCFW}. 

Briefly, if two external momenta in the amplitude $A_n$ are subjected to the on-shell BCFW shift:
\begin{align}
p_1^{\mu} \rightarrow p_1^{\mu} + z q^{\mu} \qquad p_2^{\mu} \rightarrow p_2^{\mu} - z q^{\mu}
\end{align}
and $A_n(z) \rightarrow 0$ for large $z$, then $A_n(z = 0)$ can be recursively constructed from lower-point on-shell amplitudes:
\begin{align}
\label{bcfw}
\mkern-18mu A_n=\oint \frac{dz}{z}A_n(z)=\sum_{\{L\}}\frac{A_L(\hat{1},\{L\},\hat{P})A_R(\hat{P},\{R\},\hat{n})}{P^2}. \!\!\!
\end{align}
Initial proofs required sophisticated Feynman diagram analyses, and found that gluon amplitudes have the minimum required scaling of $z^{-1}$, but that graviton amplitudes have a ``bonus", seemingly unnecessary, scaling of $z^{-2}$~\cite{BCFW}-\cite{simplestQFT}. Surprisingly, Ref. \cite{ST} found that a fully on-shell proof of BCFW constructability actually requires this improved scaling for gravitons, in order for Eq. (\ref{bcfw}) to satisfy unitarity. The bonus scaling is not just a ``bonus", but a critical property of General Relativity. 
This $z^{-2}$ scaling, also present in the case of non-adjacent gluon shifts \cite{loopBonus}, implies new residue theorems:
\begin{align}
\mkern-15mu 0=\oint A_n(z)dz=\sum_{\{L\}}z_P \frac{A_L(\hat{1},\{L\},\hat{P})A_R(\hat{P},\{R\},\hat{n})}{P^2},\!\!
\end{align}
i.e., new relations between terms in Eq.~(\ref{bcfw}): the bonus relations. The bonus scaling and the bonus relations have a number of important implications. In~\cite{non-adj}, it was shown that BCJ relations can be extracted from bonus relations. In the case of gravity, bonus relations have been used to simplify tree level calculations~\cite{bonus}. At loop level, the large $z$ scaling of the BCFW shift corresponds to the high loop momenta limit; unsurprisingly improved scaling implies improved UV behavior~\cite{simplestQFT}\cite{triangle2}\cite{Bubbles}.

In this paper, we prove that the inherent Bose-symmetry between gravitons directly implies this improved bonus scaling, completing the arguments of Ref.~\cite{ST}. Bose-symmetry in General Relativity endows it with a purely on-shell description and constrains its UV divergences\footnote{The better than expected UV behavior was also at least partially understood from the ``no-triangle" hypothesis of ${\cal N}=8$ supergravity, as a consequence of crossing symmetry and the colorless nature of gravitons in Ref.~\cite{triangle}}. We further apply the same argument to gauge theories and gravity in various dimensions.

{\bf Completing on-shell constructability.} 
Reference~\cite{ST} first assumes $n$-point and lower amplitudes scale as $z^{-1}$---thereby ensuring Eq.~(\ref{bcfw}) holds---and then checks if the BCFW expansion of the $(n+1)$-point amplitude factorizes correctly on all channels. Factorization on all channels is taken to define the amplitude. Correct factorization in most channels requires $z^{-1}$ scaling of lower point amplitudes. However, some channels do not factor correctly without improved $z^{-2}$ scaling, as well as a $ z^6$ scaling on the ``bad" shifts. In the following, we present a proof for both of these scalings.

Essentially, the argument rests on a very simple observation: any symmetric function $f(i,j)$, under deformations $i\rightarrow i+z k$, $j\rightarrow j- z k$, must scale as an even power of $z$. In particular, any function with a strictly better than $\mathcal{O}(1)$ large $z$ behavior (no poles at infinity), is automatically guaranteed to decay at least as $z^{-2}$.

Although straightforward, this is not manifest when constructing the amplitude. BCFW terms typically scale as $ z^{-1}$, but only specific pairs have canceling leading $ z^{-1}$ pieces. Similarly, the bad BCFW shift behavior of $z^6$ is only obtained when the leading $z^7$ pieces cancel in pairs. 

Consider the five point amplitude in $\mathcal{N}=8$ SUGRA, exposed by the $[1,5\rangle$  BCFW shift, where $|1] \rightarrow |1] -z |5]$ and $|n \rangle \rightarrow |n\rangle  +z |1\rangle $,
\begin{align}
M_5=&M(123P)\times M(P45)/P_{123}^2+(4\leftrightarrow3)+(4\leftrightarrow2)\nonumber \\
=&\frac{[23][45]}{\langle 12\rangle \langle 13\rangle \langle 23\rangle \langle 24\rangle \langle 34\rangle \langle 45\rangle \langle 15\rangle^2}\nonumber \\
&+\frac{[24][35]}{\langle 12\rangle \langle 14\rangle \langle 24\rangle \langle 23\rangle \langle 43\rangle \langle 35\rangle \langle 15\rangle^2}\nonumber\\
&+\frac{[43][25]}{\langle 14\rangle \langle 13\rangle \langle 43\rangle \langle 42\rangle \langle 32\rangle \langle 25\rangle \langle 15\rangle^2}  ,
\end{align}
with the SUSY-conserving delta-function stripped out. 

Under a $[2,3\rangle$ shift, the first term scales as $z^{-2}$, while the other two scale as $z^{-1}$. However, their sum (now symmetric in 2 and 3) scales as $z^{-2}$: the whole amplitude has the correct scaling. This pattern holds true in general. Where present, $ z^{-1}$'s cancel between pairs of  BCFW terms, $M_L(K_L,i,P)\times M_R(-P,j,K_R)$ and $M_L(K_L,j,P)\times M_R(-P,i,K_R)$. Further, terms without pairs over saturate the bonus scaling.

One such example is $M(1^-2^-3^+P^+)M(P^-4^-5^+6^+)$, appearing in $M_6^{\textrm{NMHV}}$. Under a $[4, 3\rangle$ shift, it has no corresponding pair: $M(1^-2^-4^-P)M(P\,3^+5^+6^+)$ vanishes for all helicities $h_P$. Luckily, it turns out these types of terms have a surprisingly improved scaling of $z^{-9}$. Hence, they never spoil the scaling of the full amplitude.

In the next section we classify and prove the scalings of all possible BCFW terms. Following this, we demonstrate how leading $z$ pieces cancel between BCFW terms.


{\bf BCFW terms under secondary $z$-shifts.} Consider the $[1,n\rangle$ BCFW expansion of a $n$-point GR tree amplitude $M_n$ (where $\tilde{\lambda}_1\rightarrow \tilde{\lambda}_1 -w \tilde{\lambda}_n$, $\lambda_n\rightarrow \lambda_n +w \lambda_1$):
\begin{eqnarray}
M_n=\sum_{L,R}\frac{M_L(\hat{1},\{L\},\hat{P})M_R(-\hat{P},\{R\},\hat{n})}{P^2} \label{1nExpand}
\end{eqnarray}
We would like to understand how  BCFW terms in $\mathcal{M}_n$ scale under secondary $[i,j\rangle$ $z$-shifts
\begin{eqnarray}
\tilde{\lambda}_i \rightarrow \tilde{\lambda}_i- z \tilde{\lambda}_j  \qquad \qquad \lambda_j \rightarrow \lambda_j + z \lambda_i . \label{shift}
\end{eqnarray}
We recall two features of these terms as they appear in Eq.~(\ref{1nExpand}). First, the value of the primary deformation parameter $w = w_P$, which accesses a given term, is
\begin{align}
w_P = \frac{P^2}{\langle 1 | P | n]} , \label{zP}
\end{align}
and, on this pole, the intermediate propagator factorizes:
\begin{align}
\hat{P}^{\alpha\dot{\alpha}}
= \frac{\{ [\tilde{\lambda}_n | P \}^\alpha \,\, \{\langle \lambda_1 | P\}^{\dot{\alpha}} }{\langle \lambda_1 | P | \tilde{\lambda}_n ]}
= \frac{\big|\lambda_P\big\rangle \big[\tilde{\lambda}_P\big|}{\langle 1|P|n]} \equiv |\hat{P} \rangle [\hat{P} |.
\end{align} 
The little-group ambiguity amounts to associating the denominator with either $\lambda_P$, $\tilde{\lambda}_P$, or some combination of them. In what follows, we find it easiest to associate it entirely with the anti-holomorphic spinor, $| \hat{P} ] = |\tilde{\lambda}_P ] / \langle 1 | P |n ]$---see Eq.~(\ref{wShift}), below.

With this in hand, we now turn to the large $z$ scalings of the various  BCFW terms, subjected to the secondary $z$-shifts in Eq.~(\ref{shift}). There will be two different types of BCFW terms: those with both $i$ and $j$ within the same subamplitude, and those with $i$ and $j$ separated by the propagator. The former inherit all $z$ dependence from the lower point amplitudes in the theory, since the secondary shift acts like a usual BCFW shift on the subamplitude. The latter are more complicated, since the $z$ shift affects the subamplitudes in several ways besides the simple shifts on $i$ and $j$.

Specifically, both $w_P$ and the factorized form of the internal propagator acquire $z$ dependence:
\begin{align}
w_P&= \frac{P^2}{\langle 1 | P | n]} 
&\longrightarrow&  \quad\quad \frac{P^2 + z \langle i | P | j]}{\langle 1 | P | n ] + z \langle 1 i \rangle [j n]} , 
\nonumber\\ 
\big| \widehat{P} \big\rangle^{\alpha}& \equiv \frac{\big| \big( [n| P \big) \big\rangle^{\alpha}}{[jn]} 
&\longrightarrow & \quad\quad \big| \widehat{P} \big\rangle^{\alpha} + z \big| i \big\rangle^{\alpha} \, , \label{wShift} \\ 
\big| \widehat{P} \big]^{\dot{\alpha}}& \equiv \frac{\big| \big(\langle 1 | P\big) \big]^{\dot{\alpha}}}{\langle 1 | P | n]/[jn] } 
&\longrightarrow& \quad\quad \frac{\big| \tilde{\lambda}_P \big]^{\dot{\alpha}} - z \langle 1 i \rangle | j ]^{\dot{\alpha}}}{\langle 1 | P | n]/[j n] - z \langle 1 i \rangle} . \nonumber
\end{align}
With this factorized form of the propagator, it turns out that the left- and right-hand subamplitudes have well defined individual $z$ scalings, which depend only on the helicity choices for $i^{h_i}$, $j^{h_j}$ and $P^h$:
\begin{eqnarray}
\label{all.scalings}
&&M_L( i^-P^-)   \sim z^{-2} 	\quad\quad M_R( j^-P^-)  \sim z^{+2}		\nonumber\\
&&M_L( i^-P^+)  \sim z^{-2}	\quad\quad M_R( j^-P^+) \sim z^{+2}		\\
&&M_L( i^+P^-)  \sim z^{+6}	\quad\quad  M_R (j^+P^-)\sim z^{+2}		\nonumber\\
&&M_L( i^+P^+) \sim z^{-2}	\quad\quad M_R (j^+P^+)\sim z^{-6} 	.	\nonumber
\end{eqnarray}
The scaling of a full BCFW term $M_LM_R/P^2$ can then be easily determined from these values, which we prove in two steps. 

First, note that the large $z$ scalings on the left of Eq.~(\ref{all.scalings}) match the familiar BCFW scalings of full amplitudes. We prove this by showing that the large $z$ behavior of the left-hand subamplitude maps isomorphically onto a  BCFW shift of $M_L$. Looking at Eq.~(\ref{wShift}), we see that, in the large $z$ limit, the spinors of $i$ and $P$ become
\begin{align}\label{miracle}
\lambda_i &\longrightarrow \lambda_i&\lambda_P&\longrightarrow z \lambda_i\nonumber \\
 \tilde{\lambda}_i &\longrightarrow - z \tilde{\lambda}_j &  \tilde{\lambda}_P&\longrightarrow \tilde{\lambda}_j,
\end{align}
which is just a regular BCFW $[i,P\rangle$ shift within the left-hand subamplitude. 

Now we turn to the slightly unusual scalings on the right-hand side of Eq.~(\ref{all.scalings}). With the little-group choice in Eq.~(\ref{wShift}), the left-hand subamplitude has exactly the correct spinor variables to map onto the usual  BCFW shift. Now observe that, starting with the other little-group choice for the spinors on the $z$ shifted internal propagator, we obtain the usual BCFW scalings on this side:
\begin{eqnarray}
\label{right.good}
&&M_R( j^-P^-)\sim z^{-2}\nonumber\\ 
&&M_R( j^-P^+)\sim z^{+6} \nonumber \\
&&M_R (j^+P^-)\sim z^{-2}\nonumber \\
&&M_R (j^+P^+)\sim z^{-2} . 
\end{eqnarray}
Proving these results is identical to the previous reasoning for the left-hand subamplitude. 

It becomes clear now that to get the other half of the scalings, we need only account for the change in $z$ scaling when switching the $1/\langle 1 | P(z) | n]$ factor between $\lambda_P$ and $\tilde{\lambda}_P$. Assume that the spinors of the propagator appear with weights\footnote{In general, the spinors need not appear with uniform homogeneity. The analysis below still holds, but must be applied term by term. The same caveat applies to Eqs. (\ref{mij}) and (\ref{mij2}).}:
\begin{equation}
\label{propto}
M_R \propto \big(|P\rangle\big)^a \big(|P]\big)^b  , 
\end{equation}
where $-a+b=2 h_P$, and $h_P$ is the helicity of the internal propagator as it enters the right-hand subamplitude. Now, in the limiting cases where $1/\langle 1 | P(z) |n]$ is entirely associated with $|\lambda_P\rangle$ or $|\tilde{\lambda}_P]$ the amplitude scales as:
\begin{align}
&M_R\propto \bigg(\frac{|\lambda_P\rangle}{\langle 1|P|n]}\bigg)^a\big(|\tilde{\lambda}_P]\big)^b \rightarrow z^s, \quad {\rm or}\\
&M_R\propto \big(|\lambda_P\rangle\big)^a\bigg(\frac{|\tilde{\lambda}_P]}{\langle 1|P|n]}\bigg)^b \rightarrow z^t,
\end{align}
where $s$ is the BCFW large $z$ scaling exponent, obtained in Eq.~(\ref{right.good}), and $t$ is the related scaling, for the other internal little-group choice. It follows that $s-t=b-a= 2h_P$, and so the $t$ scalings can be easily derived as $t=s\pm 4$, depending on the helicity of the propagator.

Having proven all eight scaling relations in Eq.~(\ref{all.scalings}), we can classify the scaling behavior of all possible types of  BCF terms with $i$ and $j$ in different subamplitudes. For these terms the propagator contributes a $z^{-1}$ to each term, and so from Eq.~(\ref{all.scalings}) we obtain eight possible types of terms: 
\begin{align}
&\bullet\ M_L( i^+P^-) M_R( j^-P^+)/P^2 \,\, \textrm{scales as $ z^{+7}$},\label{bd}\\
&\bullet\ M_L( i^-P^-) M_R( j^+P^+)/P^2 \,\, \textrm{scales as $ z^{-9}$},\label{sgd}\\ 
&\bullet\ \textrm{The other six  BCFW terms scale as $ z^{-1}$ }.\label{gd}
\end{align}
In the next section we will see how pairing terms improves these scalings by one power of $z$, such that we recover the required $z^{-2}$ and $z^6$ scalings. 

Finally, while the individual scalings in Eq.~(\ref{all.scalings}) are not invariant under $z$ dependent little-group rescalings on the internal line $\widehat{P}(z)$, the above results for full BCFW terms are invariant under these rescalings.


{\bf Improved behavior from symmetric sums.} 
We first study $[+,+\rangle$ and $[-,-\rangle$ shifts, with scalings in Eq.~(\ref{gd}). Define $M_L(K_L,i,P)\times M_R(-P,j,K_R)/P^2\equiv M(i|j)$, where $K_L$ is the momenta from the other external states on the left-hand subamplitude. We wish to show that in the large $z$ limit
\eq
M( i | j )=-M( j | i )  .
\eqe
so the leading $z^{-1}$ pieces cancel in the symmetric sum of BCFW terms, $M( i | j )+M( j | i )$. 

Because $i$ and $j$ have the same helicity, $M( j | i )$ is obtained directly from $M( i | j )$ by simply swapping labels:
\begin{align}
M(i|j)=M(\lambda_i,\tilde{\lambda}_i, \lambda_j, \tilde{\lambda}_j)\\
M(j|i)=M(\lambda_j,\tilde{\lambda}_j, \lambda_i, \tilde{\lambda}_i)
\end{align}
In the large $z$ limit, these become
\begin{align}
M(i|j)=M(\lambda_i,-z\tilde{\lambda}_j, z\lambda_i, \tilde{\lambda}_j)\\
M(i|j)=M(z\lambda_i,\tilde{\lambda}_j, \lambda_i, -z\tilde{\lambda}_j)
\end{align}
The two have equal $z$ scaling, and so can only differ by a relative sign. The spinors appear with weights
\begin{align}\label{mij}
&M( i | j )\propto \langle ij\rangle^F [ij]^G (\lambda_i)^a(\tilde{\lambda}_i)^b (\lambda_j)^c(\tilde{\lambda}_j)^d\nonumber \\
&M( j | i )\propto \langle ji\rangle^F [ji]^G (\lambda_j)^a(\tilde{\lambda}_j)^b (\lambda_i)^c(\tilde{\lambda}_i)^d  ,
\end{align}
while in the large $z$ limit, the leading terms are
\begin{align}
M( i | j )&\propto z^{b+c}\left(\langle ij\rangle^F [ij]^G (\lambda_i)^a(-\tilde{\lambda}_j)^b (\lambda_i)^c(\tilde{\lambda}_j)^d \right)\nonumber\\
M( j | i )&\propto z^{a+d}\left(\langle ji\rangle^F[ji]^G (\lambda_i)^a(\tilde{\lambda}_j)^b (\lambda_i)^c(-\tilde{\lambda}_j)^d \right)\!.
\end{align}
These cancel if and only if $F+G+b+d=\textrm{odd}$. First, from Eq.~(\ref{gd}), $M( a | b )$'s scale as $z^{\rm odd}$. So $b + c = a + d = {\rm odd}$. Second, by helicity counting in Eq. (\ref{mij}), we know $-F+G - c +d = 2 h_j  = {\rm even}$. Therefore, we obtain the required result, and the leading $z^{-1}$ pieces cancel.

For the $[-,+\rangle$ and $[+,-\rangle$ shifts a simple modification of the above argument is required. This is because we now expect the cancellation to occur between the pair terms $M_L(K_L,i^-,P^+)\times M_R(-P^-,j^+,K_R)/P^2$ and $M_L(K_L,j^+,P^-)\times M_R(-P^+,i^-,K_R)/P^2$. Switching different helicity particles requires us to flip the propagator's helicity as well. It can be shown that, in the large $z$-limit, $M_L(K_L,i^-,P^+)= M_L(K_L,j^+,P^-)$; likewise for the right-hand subamplitude. Note that switching $i^-$ and $j^+$ requires more care now: functionally, the correct label swaps for $M_L$ are $i \rightarrow P$, $P \rightarrow j$ while for $M_R$ $j \rightarrow P$ and $P \rightarrow i$. Therefore we can write, as above,
\begin{align}\nonumber\label{mij2}
M_L(i^-,P^+)
 &\propto  \langle iP\rangle^F [iP]^G  (\lambda_i)^a(\tilde{\lambda}_i)^b (\lambda_P)^k (\tilde{\lambda}_P)^l\\
M_L(j^+,P^-)
 &\propto  \langle Pj\rangle^F [Pj]^G  (\lambda_P)^a(\tilde{\lambda}_P)^b (\lambda_j)^k (\tilde{\lambda}_j)^l .
\end{align}
Crucially, the large $z$ limit is also different for the two subamplitudes, since the limits (\ref{miracle}) were obtained with $i\in P$. The second subamplitude instead has $j\in P$, and in this case the limits are $\lambda_P \rightarrow -z \lambda_i$ and $\tilde{\lambda}_P \rightarrow \tilde{\lambda}_j$. In the large $z$ limit then identical counting as above shows that $a+b=\rm even$, and the same will hold for $M_R$. The propagator is antisymmetric in the large $z$ limit under swapping $i$ and $j$, and therefore the leading $z$ pieces cancel as expected. This cancellation reduces the leading $z^{-1}$ and $z^{+7}$ scalings for the opposite helicity shifted  BCF terms in the previous section, down to the well known $z^{-2}$ and $z^{+6}$  BCFW scalings for GR. This completes the proof of the bonus scaling for GR, and closes the final gap in the on-shell proof of BCFW in GR Ref.~\cite{ST}.


{\bf Analysis of the full amplitude.}
The simple argument we used above can be applied directly to the whole amplitude, if we restrict to like-helicity shifts. Consider
\begin{align}
\!\!\!\!\!
A_n(i,j)\propto \langle ij\rangle^F [ij]^G  (\lambda_i)^a( \tilde{\lambda}_i)^b  (\lambda_j)^c(\tilde{\lambda}_j)^d .\label{Amp}
\end{align}
If this amplitude is manifestly symmetric under exchange of two (bosonic) particle labels, then $A_n(i,j) = A_n(j,i)$, which fixes $a = c$, $ b = d$, and $F+G = \rm even$. By helicity counting, $-F+G-a+b = 2h_i=\rm even$, and then $a + b=\rm even$. So, under a $[i,j\rangle$ shift,
\eq
A_n(i(z),j(z)) \sim z^{b+c}=z^{a+b} = z^{\rm even} . \label{ScaleFin}
\eqe

This same logic holds in Eq.~(\ref{Amp}), even if the shifted lines are identical fermions. Permuting labels $i$ and $j$ again forces $a = c$, and $b = d$, and $F+G=\rm odd$. But so must $2h_i = -F +G - a + b$. Hence $a + b$ remains even.  BCFW shifts of identical particles, bosons or fermions, fix $z^{\rm even}$ scaling at large $z$.

To understand the opposite-helicity shifts, we are led to consider pure GR as embedded within maximal ${\cal N}=8$ SUGRA. Amplitudes in maximal supergravity do not distinguish between positive and negative helicity graviton states. Using the methods of \cite{nLess} to truncate to pure GR, we recover the usual BCFW scalings.

As an interesting corollary of our four-dimensional analysis, the large $z$ scaling of gravity amplitudes in three dimensions is drastically improved to $z^{-4}$. Due to the fact that the little group in three dimensions is a discrete group, the BCFW deformation is non-linear. In particular the three dimensional spinors shift as~\cite{3d}:
\begin{align}\label{What}
\!\!\!\!\!\!
\lambda_i(z) \! = \! {\rm ch}(z) \lambda_i \! + \! {\rm sh}(z) \lambda_j \, , \, 
\lambda_j(z) \! = \! {\rm sh}(z) \lambda_i \! + \! {\rm ch}(z) \lambda_j \!\!\!\!\!\!
\end{align}
where ${\rm ch}(z) = (z + z^{-1})/2$ and ${\rm sh}(z) = (z - z^{-1})/2i$. Thus, momenta shift as
\begin{align}\label{What2}
p_i(z) = \overline{P_{ij}} + y q + \frac{1}{y}\tilde{q} \,\,\, , \,\,\, 
p_j(z) = \overline{P_{ij}} -  y q  - \frac{1}{y}\tilde{q}
\end{align}
where $\overline{P_{ij}} = \frac{p_i+p_j}{2}$, $y=z^2$, and $q,\tilde{q}$ can be read off from Eq.~(\ref{What}). Now let's consider three-dimensional gravity amplitudes that arise from the dimension reduction of four-dimensional gravity theory. The degrees of freedom are given by a dilaton and a scalar. Since both are bosons, little group dictates that one must have even power of $\lambda_i$. Thus the large $z$ behavior of gravity amplitudes is completely dictated by Eq.~(\ref{What2}). Permutation invariance then requires the function to be symmetric under $y\leftrightarrow -y$, and so must be an even power of $y$. Thus if gravity amplitudes can be constructed via BCFW shift, the large $z$ asymptotic behavior must be at most $y^{-2}=z^{-4}$. Indeed it is straightforward to check that the four-point $\mathcal{N}=16$ supergravity amplitude behaves as $z^{-4}$ under a super-BCFW shift. This is to be compared with the $z^{-1}$ scaling of superconformal Chern-Simons theory~\cite{3d}.

More generally, BCFW shifts in $d \geq 4$ take the form,
\begin{align}
&&p_i^{\mu}(z) = p_i^{\mu} + z \, q^{\mu} \qquad p_j^{\mu}(z) = p_j^{\mu} - z \, q^{\mu}  ,
\end{align}
where $q$ is null and orthogonal to $p_i$ and to $p_j$. External wave-functions of shifted boson lines also shift~\cite{SpinLorentz}. For identical bosons, Bose-symmetry disallows $z^{\rm odd}$ scaling, as it would introduce a sign change under label swaps. Identical fermions shift similarly; here the antisymmetric contraction of the identical spinor wave-functions absorbs their exchange-sign.  BCFW shifts of identical particles must scale as $z^{\rm even}$ for large $z$ in dimensions $d \geq 4$.

Symmetry between identical particles is crucial for these cancellations to occur. Gluon partial amplitudes are not permutation invariant: distinct gluons generally have different colors. This spoils the permutation invariance---as is clear from $ z^{-1}$ drop-off of adjacent shifts of a color-ordered tree amplitude in Yang-Mills. Gravitons, however, are unique: they cannot have different ``colors'' \cite{WW}. Thus graviton amplitudes are invariant under permutations from the outset: the discrete symmetry group of graviton amplitudes is larger than for gluon amplitudes. Consequently, gravity amplitudes are softer in the deep-UV than Yang-Mills amplitudes. 


{\bf Bose-symmetry and color in Yang-Mills.}
Finally, we explore the interplay between color and the large $z$ structure of Yang-Mills amplitudes. For ease, we focus on $A^{\rm tree}_4(1^-,2^-,3^+,4^+)$. It can be written in terms of color-ordered partial amplitudes as
\eq
\!\!\!\! \frac{A_4(1^-2^-3^+4^+)}{\langle 12\rangle^2 [34]^2 } = \frac{Tr(1234)}{st} + \frac{Tr(1243)}{su} + \frac{Tr(1324)}{tu}. \label{YMamp}
\eqe
Under a $[1,2\rangle$ shift, only $t$ and $u$ shift, and in opposite directions: $\widehat{t}(z) = t + z \langle 1 | 4 | 2]$, and $\widehat{u}(z) = u - z  \langle 1 | 4 | 2]$. The term proportional to $Tr(1324)$ scales as $z^{-2}$, while the other two scale as $ z^{-1}$. The leading $z$ terms,
\eq
\frac{A_4(\widehat{1}^-,\widehat{2}^-,3^+,4^+)}{\langle 12\rangle^2 [34]^2 } \sim \frac{Tr(1234)-Tr(1243)}{z \langle 1 | 4 | 2] \, s} + \cdots , \label{YMpair}
\eqe
cancel when gluons 1 and 2 are identical, and $T_1 = T_2$. 

Cancellation of $ z^{-1}$ terms must hold for general tree amplitudes when the gluons have the same color labels. However, only  BCFW shifts of lines that are \emph{adjacent} in color-ordering cancel pairwise as in Eq.~(\ref{YMpair}). For color-orderings where this shift is non-adjacent, there are no pairs of  BCF terms with canceling $ z^{-1}$-terms. This implies that good non-adjacent  BCFW shifts in gluon partial amplitudes must scale as $z^{-2}$.


{\bf Future directions and concluding remarks.}
We have shown the $ z^{-2}$ bonus scalings/relations, crucial for consistent on-shell contraction of Gravitational S-matrices, follow from Bose-symmetry. Similar $ z^{-1}$ cancellations occur in QED and GR~\cite{photon}. Further, Bose-symmetry alone implies $ z^{-2}$ drop-off of non-adjacent  BCFW shifts in Yang-Mills. More broadly, BCFW shifts of identical particles---bosons and fermions---must scale as $z^{\rm even}$ in general settings, beyond $d=4$.

Graviton amplitudes in Refs. ~\cite{MHVtree}\cite{DAM}\cite{Hodges2012}, which manifest permutation symmetry, also manifest $ z^{-2}$ drop-off. This is not a coincidence: permutation symmetry automatically implies bonus behavior. A better understanding of gravity should be tied to more natural manifestations of permutation invariance. However, not all improved scalings obviously come from permutation invariance. Notably, Hodges' observation that BCFW-terms, built from ``bad'' ``opposite helicity'' $z^{-1}$ $\mathcal{N}=7$ SUGRA shifts, term-by-term scale as $z^{-2}$~\cite{Hodges2011}. As the legs are not identical, permutation invariance is not prominent in the proof~\cite{Kickass}. 

Permutation invariance has unrecognized and powerful consequences even at tree level. Do new constraints appear when accounting for it in other shifts? Does it have non-trivial consequences at high-loop orders in ${\cal N}=8$ SUGRA, or ${\cal N} = 4$ SYM? Would mandating it expose new facets of the ``Amplituhedron'' of Ref.~\cite{SYM}?

{\bf Acknowledgements:} We thank Aleksey Cherman, Jaroslav Trnka and Song He for careful readings and comments. We specially thank Yu-tin Huang for thoughtful conversations, especially regarding color in QCD, the discussion on $z^{-4}$ scaling in 3d gravity and generalizing Eq.~(\ref{ScaleFin}) to arbitrary dimension, and Nima Arkani-Hamed for discussions and encouraging us to clarify the scalings of individual BCFW terms.


\end{document}